\documentstyle[aps,multicol,pre]{revtex}

\begin{document}
\title{A model for complex aftershock sequences}
\author{Y. Moreno,$^{1,4}$ A. M. Correig,$^2$ J. B.
G\'{o}mez,$^3$ A. F. Pacheco,$^1$}
\address{
$^1$ Departamento de F\'{\i}sica Te\'{o}rica, Universidad de Zaragoza, 50009
Zaragoza, Spain.\\
$^2$ Departament d'Astronomia i Meteorologia, Universitat de Barcelona, E-08028 Barcelona, Spain.\\
$^3$ Departamento de Ciencias de la Tierra, Universidad de Zaragoza, 50009
Zaragoza, Spain.\\
$^4$ Now at The Abdus Salam International Centre for
Theoretical Physics, Condensed Matter Group, P.O. Box 586, Strada
Costiera 11, I-34100, Trieste, Italy.}
\date{\today}
\maketitle
\widetext
\begin{abstract}
The decay rate of aftershocks is commonly very well described by
the modified Omori law, $n(t) \propto t^{-p}$, where $n(t)$ is the
number of aftershocks per unit time, $t$ is the time after the
main shock, and $p$ is a constant in the range $0.9<p<1.5$, and
usually close to 1. But there are also more complex aftershock
sequences for which the Omori law can be considered only as a
first approximation. One of these complex aftershock sequences
took place in the Eastern Pyrenees on February 18, 1996, and was
described in detail by {\it Correig et al.} [1997]. In this paper,
we propose a new model inspired by dynamic fiber-bundle models to
interpret this type of complex aftershock sequences with sudden
increases in the rate of aftershock production not directly
related to the magnitude of the aftershocks (as in the
epidemic-type aftershock sequences). The model is a simple,
discrete, stochastic fracture model where the elements (asperities
or barriers) break because of static fatigue, transfer stress
according to a local load-sharing rule and then are regenerated.
We find a very good agreement between the model and the Eastern
Pyrenees aftershock sequence and we propose that the key mechanism
for explaining aftershocks, apart from a time-dependent rock
strength, is the presence of dynamic stress fluctuations which
constantly reset the initial conditions for the next aftershock in
the sequence.
\end{abstract}
\pacs{PACS number(s):}
\begin{multicols}{2}
\narrowtext

\section{Introduction}

Omori discovered scaling in earthquakes in the frequency distribution
of aftershocks over one hundred years ago when he proposed a formula to
represent the decay of aftershock activity with time [{\it
Omori,} 1894]. Now, one hundred years later, it remains as one of the
few well established empirical laws in Seismology. As noted by {\it
Utsu} [1995], `any theory for the origin of aftershocks must explain
this law, which is unique for its power law dependence on time'. The
Omori law (as modified by {\it Utsu}, 1961),

\begin{equation}
n(t) = Kt^{-p},
\end{equation}
says that the number of aftershocks $n(t)$, measured at time $t$
after the time of the main shock, declines following a power law
with an exponent $p$ around one ($0.9 < p < 1.5$, with a median of
about 1.1, [{\it Utsu,} 1995]), and $K$ being a
proportionality constant. To avoid divergence at $t=0$, the Omori
law is usually written in the form

\begin{equation}
n(t) = K(t+c)^{-p},
 \label{eq4002}
\end{equation}
where $c$ is an additional `small' positive constant with
dimensions of time (between 0.01 and 1 days, with a median of 0.3
days, [{\it Utsu,} 1995]) . The power law, scale-free
behavior is maintained for $t\gg c$, with a transition to
$n(t)=\mbox{const}$ for $t\le c$ that accounts for incompleteness
in the detection of low-magnitude aftershocks during the few hours
after the mainshock (e.g. [{\it Utsu,} 1995, fig 2;
{\it Gross and Kisslinger} 1994]. From Eq.\
(\ref{eq4002}), the cumulative number of aftershocks, $N(t)$,
occurred until time $t$ after the main shock, defined as $\int^t_0
n(s)ds$, is
\begin{equation}
N(t)= \left\{
\begin{array}{ll}  K \ln (t/c+1) & \mbox{if }p=1 \\
K\lbrace c^{1-p}-(t+c)^{1-p} \rbrace/(p-1) & \mbox{if }p\ne1
\end{array} \right.
\end{equation}

The Omori law has also been verified in laboratory-scale
experiments in brittle rock deformation by measuring acoustic
emission [{\it Scholz,} 1968a,b; {\it Lockner
and Byerlee}, 1977; {\it Hirata,} 1987; {\it Sammonds et al.,} 1992; {\it Lockner,} 1993] and in
mine-induced seismicity [{\it Talabi,} 1997], which
represents an intermediate scale between lab experiments and
natural aftershocks. The fulfilment of the Omori law at the
microscale ($10^{-3}-10^{-1}$ m), mesoscale ($1-10^{1}$ m) and
macroscale ($10^{2}-10^{4}$ m), suggests that a universal process
is behind the inelastic strain responsible for acoustic emission
in the laboratory [{\it Hirata,} 1987], induced
microseismicity in mines [{\it Gibowicz,} 1997] and
aftershock sequences in active tectonic faults [{\it
Utsu,} 1995; {\it Gross and Kisslinger} 1994]. But,
what is this mechanism?

Static fatigue, also known as stress-, creep-, or delayed fracture
is the basic way of time-dependent failure under a constant load
of a broad variety of materials, including textile fibers
[{\it Coleman,} 1957], fiber composites [{\it
Phoenix,} 1977], wood [{\it Garcimart\'{\i}n et al.,}
1997], microcrystals [{\it Pauchard and Meunier,}
1993], gels [{\it Bonn et al.,} 1998], policrystalline
ceramics [{\it Jacobs and Chen,} 1994], metals
[{\it Schleinkofer et al.,} 1996], silicate glasses
[{\it Charles,} 1958], minerals [{\it
Scholz,} 1968a; {\it Barnett and Kerrich,} 1980], and rocks
[{\it Atkinson,} 1984]. In all these cases, the
signature of static fatigue is the observation of a failure
strength that is a function of the load history of the material.
For brittle materials, and from the point of view of fracture
mechanics, time-dependent strength is commonly associated with
{\it kinetic fracture}, {\it i.e.}, with the propagation of cracks
under a crack tip stress intensity factor below the modulus of
cohesion of the material [{\it Kostrov et al.,} 1969].
This propagation is stable and quasi-static, and is referred to as
subcritical crack growth, where `quasi-static' means at velocities
much less than the sonic velocity of the medium [{\it
Das and Scholz ,} 1991]. The presence of a chemically active
fluid environment saturating the pore and crack space enhances
this subcritical crack growth, a mechanism known as {\it stress
corrosion} [{\it Charles,} 1958; {\it Wiederhorn,}
1967]. There is ample evidence that geological materials under
brittle conditions owe their time-dependent strength to the
mechanism of subcritical crack growth assisted by stress corrosion
[{\it Atkinson,} 1984; {\it Atkinson and Meredith,}
1987].

{\it Benioff} [1951] presented the first detailed theory offering an
explanation of the causes and characteristics of aftershock sequences
in terms of identifiable mechanical properties. According to his
theory, aftershocks occur when there is a {\it time-dependent} recovery
of stress following the main shock. The stress recovery was ascribed by
Benioff to a static fatigue of the rocks in the immediate area of the
fault.

Since this seminal paper, many laboratory and numerical experiments
have confirmed the hypothesis that aftershocks are a process of
relaxing stress concentrations produced by the dynamic rupture of the
main shock, and that they are, therefore, an intrinsic time-dependent
rheological effect. In this context, {\it Scholz} [1968b] formulated
the first time-dependent strength model of aftershocks. He suggested
that a time-dependent strength of the rocks in the area of the main
shock could be the cause of the aftershock sequences and invoked static
fatigue due to local overloads to stresses much higher than their
long-term strength as the main mechanism of aftershocks. Based on {\it
Scholz}'s [1968a] laboratory experiments on static fatigue of quartz,
{\it Das and Scholz} [1981] formulated a general model of aftershocks
using elastic fracture mechanics and the concept of subcritical crack
growth. They showed that this model is consistent with the decay rate
of aftershocks as expressed by the Omori law, and that it is able to
reproduce many other characteristics of real aftershock sequences. More
recent works and papers that stress the role of time-dependent strength
in aftershock dynamics are: {\it Yamashita and Knopoff} [1987], who
assume that stress corrosion is the physical mechanism for the delayed
fracture in aftershocks; {\it Marcellini} [1995, 1997], advocating
static fatigue, together with stress inhomogeneities, as the cause of
Omori-law aftershock sequences; and {\it Lee} [1999] and {\it Lee and
Sornette} [1999], who constructed a fuse network model of aftershocks
incorporating a time dependent strength compatible with the mechanism
of subcritical crack growth. All these models of aftershocks also obey
the Omori law.

As mentioned above, the Omori-law decay rate of aftershocks
following a main shock is an almost universal characteristic of
seismicity (as compared to the more irregular patterns of
premonitory activity as foreshocks or quiescence). But despite
this universality, many real aftershock sequences display
anomalies in the decay rate that depart from the simple Omori-law
behavior. Among these anomalies we can cite [{\it Utsu,}
1995]: (i) cases in which seismic activity following the main
shock cannot be represented by a simple power law due to the
mixing of different series of activity [{\it Gross and
Kisslinger} 1994]; and (ii) cases where aftershocks decay, as a
whole, according to the Omori law, but depart temporarily from the
formula due to abrupt changes in activity (accelerations and/or
quiescence).

In this paper we are interested in aftershock series that do not
rigorously follow the Omori law and, in particular, in this second
type of anomalies where sudden accelerations in the rate of
aftershock activity are not directly linked to aftershocks of
larger magnitude. This last case is the so-called epidemic-type
aftershock sequence, ETAS, where each aftershock has its own
sequence of aftershocks [{\it Ogata,} 1988], and can be
thought of as a fractal version of the simple Omori relaxation
formula. There are, however, some aftershock sequences where the
changes in decay rate are {\it independent} of the magnitude of
the aftershocks that provoke these changes in activity, and that
cannot be ascribed to the ETAS model. One of these aftershock
sequences took place in the eastern Pyrenees on February 18, 1996
[{\it Correig et al.,} 1997] and in this paper we
propose a simple stochastic model {\it \'a la} dynamic fiber
bundle model as a framework to interpret this class of aftershock
sequences.

\section{The data}

On February 18, 1996, a local magnitude $M_L=5.2$ earthquake
occurred in the Eastern Pyrenees, with epicentral location
N42$\deg$47.71', E2$\deg$32.30' and focal depth of 8 km
[{\it Rigo et al.,} 1997; {\it Pauchet et al.,} 1999].

The series of aftershocks that followed this event was recorded at
the three-component continuous broadband seismic station at the
Tunel del Cad\'{\i}, located at about 80 km SW of the epicentral
area. Altogether, the series consists of 337 events (complete for
a threshold magnitude of 1.9), spanning 1846 hours (77 days) from
the time of the main shock, and with magnitudes ranging from 1.9
to 3.8. Figure \ref{fig:acum}a shows the cumulative series of
aftershocks, along with the magnitude of the events. The sudden
change  in slope at about 300 hours is not due to incompleteness
of the series, and from the point of view of the magnitude of the
aftershocks, there is no specific characteristic, nor any relevant
event, that justifies this sudden change in the event rate.
Because of this different behavior, we will restrict our attention
to the series defined by the first 300 hours, with a total of 308
events, as displayed in Figure \ref{fig:acum}b.

The most striking feature of this series is the change in
concavity of the cumulative curve not correlated to any
significant event (as would be the case from the point of view of
an ETAS model), suggesting an increase in the rate of aftershocks
production apparently not related to any relaxation process.If we
try to fit by the maximum likelihood method the aftershock data in
Fig.~\ref{fig:acum}b to Eq. (3), we immediately appreciate that a
unique fit to the whole range (0-300 hours) is graphically worse
than two independent fits to the ranges 0-100 hours and 140-300
hours. Unfortunately, the parameter estimation for the 140-300
hours is not unique, in the sense that more than one set of
parameter values return the same value of the likelihood function
after maximization. After fixing the values of $K$ and $c$ (to the
values $K=17\pm 2$ and $c=0.2\pm 0.1$ obtained for the fit to the
0-100 hours interval), the 140-300 hours fit returns a unique
value for the parameter $p=0.654 \pm 0.005$, which is
statistically different to the value $p=0.75 \pm 0.04$ obtained
for the 0-100 hours interval. The error in both estimates is one
standard deviation assuming a normal distribution for the
variance. These values of $p$ are abnormally low (0.75 for the
0-100 hours interval, and 0.65 for the 140-300 hours interval).
The fit and the value of $p$ do not improve if the magnitude
threshold is increased [{\it Correig et al.,} 1997].

The interpretation of the Omori law as a relaxation process
suggests a way of separating the aftershocks in the series into
two classes: class A for the events that follow a relaxation law
and class B for those events that do not. The criterion to assign
the events to classes A or B is the following: if the interval of
time $\Delta t_i$ between events $i$ and $i-1$ is strictly larger
than the interval of time $\Delta t_{i-1}$ between events $i-1$
and $i-2$, the event $i$ belongs to class A; otherwise it belongs
to class B. Events belonging to class A are termed {\it leading
aftershocks}, and those belonging to class B, {\it cascades}.
Figure \ref{fig:acum}c shows the aftershock sequence classified as
leading events (solid circles) and cascades (dots). Note that a
cascade is initiated by a leading aftershock and that this leading
aftershock has no significative different magnitude.

The fit of Eq. (3) to the series formed by the leading aftershocks
is shown in Fig.~\ref{fig:cascades}a. The fit has improved
significantly from the initial fit to the whole sequence, and the
value obtained for the exponent is now $p=0.94$, much more in
agreement with the standard values for worldwide aftershock
sequences. Figure \ref{fig:cascades}b depicts the series of
cascades, in which the first term of each cascade is a leading
aftershock. Two important features are readily visible from the
figure: ($i$) the cascades are in general well approximated by
straight lines; and ($ii$) their corresponding slopes decrease
with time. A plot of the slope $s$ of the cascades against time
$t$ (Figure \ref{fig:slopes}) shows the remarkable fact that there
exists a power-law relationship of the form $s\propto t^{-\nu}$
between them, with $\nu \approx 0.71$.

The properties summarized in Figs.~\ref{fig:acum} through
\ref{fig:slopes} for the aftershock sequence of the Eastern
Pyrenees can be described at first order with the modified Omori
law, Eqs. (2) and (3). But at second order there are important
non-random fluctuations about this law (represented by the
cascades) that can not be fitted in detail with, nor accounted
for, the Omori law and its relaxation origin. In the next Section
we will construct a model to account for this second order
deviations from the Omori law, and for the Omori law itself, of
course.

We want to stress here that the characteristics of the series of
aftershocks from the February 18, 1996, Pyrenees mainshock are by
no means `exceptional'. On the contrary, they seem to be a rather
general feature of aftershock series. The authors are currently
analyzing various aftershock sequences (Greece, Kobe, Landers,
Northridge) and have found a behavior very similar to that of the
Pyrenees aftershock sequence. Results will be reported elsewhere.

\section{Dynamic fiber-bundle models}

Fiber-bundle models (FBMs) are simple discrete stochastic fracture
models amenable to either close analytical or fast numerical
solution.  These models arose in intimate connection with the
strength of bundles of textile fibers [{\it Daniels,}
1945; {\it Coleman,} 1957]. Since Daniels' and
Coleman's seminal works there has been a long tradition in the use
of these simple models to analyze failure of heterogeneous
materials [{\it Vazquez-Prada et al.}, 1999 and references
therein].

The dynamic version of the FBM simulates the failure of materials
because of static fatigue or delayed rupture. In this version, one
considers: ($i$) a discrete set of $N$ elements located on the
sites of a $d$-dimensional lattice; ($ii$) a probability
distribution for the nominal lifetimes of individual elements; and
($iii$) a load-transfer rule which determines how the load carried
by a failed element is to be distributed among the surviving
elements in the set.

As stated in ($ii$), the nominal lifetimes, $t_j,$ of the
individual elements supporting an initial stress $\sigma_i$, equal
for all $j$, are taken from a probability distribution of the type
\begin{equation}
n_j=1-e^{-k(\sigma_i)t_j}\quad\qquad j=1,2,\ldots,N.
\end{equation}
where $n_j$ are random numbers ($0\leq n_j\leq 1$) and $k(\sigma_i)$ is
the so called hazard rate or breaking rule. The most accepted hazard
rate is of the form
\begin{equation}
k(\sigma_i)=\nu_0 \left( \frac{\sigma_i}{\sigma_0} \right)^\rho,
\label{eq4015}
\end{equation}
$\nu_0$ and $\sigma_0$ represent a hazard rate of reference and a
stress of reference respectively, and $\rho$ is an exponent in the
range $2< \rho <50$. This function has been used to fit
experimental results of time-to-failure on various materials
[{\it Coleman,} 1957; {\it Phoenix, } 1977 ]. Besides,
{\it Phoenix and Tierney} [1983] derived it from a kinetic theory
of thermally activated atomic bond rupture [{\it
Zhurkov,} 1965], and showed that in many circumstances it is a
better approximation than the exponential breaking rule,
$\kappa(\sigma)=\alpha \exp(\beta\sigma)$, also used in modeling
time dependent fracture [{\it Coleman,} 1957].

Equation\ (\ref{eq4015}) has the same form as Charles power-law to
describe stress corrosion induced subcritical crack growth in
geological materials [{\it Atkinson,} 1984]:
\begin{equation}
v=v_0 \exp(-H/RT)K_{I}^n, \label{eq40112}
\end{equation}
where $v$ is the crack velocity, $H$ the activation energy, $R$
the gas constant, $T$ the absolute temperature, $K_{I}$ the stress
intensity factor for mode I fracture, and $v_0$ and $n$ are
constants. Sometimes, $n$ is known as the stress corrosion index.
Nominal values at room temperature and in wet rock are
[{\it Atkinson and Meredith,} 1987]: 15-40 for quartz
and quartz rocks; 10-30 for calcite rocks; 30-70 for granitic
rocks; and 25-50 for gabbro and basalt. If we assume constant
temperature, Eq.\ (\ref{eq40112}) can be simplified to
\begin{equation}
v=AK_I^n, \label{eq40113}
\end{equation}
which is identical to Eq.\ (\ref{eq4015}) if we substitute $\sigma$ by
$K$ and identify the breaking rate expressed by Eq.\ (\ref{eq4015})
with the crack opening velocity expressed by Eq.\ (\ref{eq40113}).

In dynamic FBMs, once elements begin to fail because of fatigue
and their stress is transferred according to the assumed rule, the
stresses among the surviving elements are no longer equal and the
stress history of an individual element becomes complicated by the
successive step-like transfers coming from failing elements. The
effect of the increase in stress for a particular element $j$ is
the reduction of its lifetime from the initially assigned $t_j$ to
a $T_j$ defined by
\begin{equation}
t_{j}=\int_0^{T_{j}}\left(\frac{\sigma_j(t)}{\sigma_i}\right)^\rho dt .
\label{eq406}
\end{equation}
Notice that in the case of independent elements (i.e., no stress
transfer), $\sigma_j(t)=\sigma_i$ $\forall j$ and hence the bundle
would break as a succession of individual failures at the times
$t_j$ assigned at the beginning. When, on the contrary, stress
redistribution between elements is assumed, the temporal series of
individual failures actually occurs at the times $T_j$ dictated by
Eq.\ (\ref{eq406}).

Note, that in these models, the total stress acting on the system
is conserved until the failure of the last element; and for the
reasons explained above, the last element, i.e., that with the
longest $T_j$, does not coincide, in general, with that with the
longest $t_j$. A detailed explanation of how to perform a Monte
Carlo simulation using this model can be found in [{\it
Newman et al.,} 1995] where the reader is referred to for
details.

The dynamic FBM can also be calculated using a probabilistic
approach. This approach was introduced by us in {\it
G\'{o}mez et al.,} [1998]. From this perspective, one starts
with $N$ elements loaded with an initial common stress equal to
$\sigma_i$. The mean time interval, $\delta$, for one element to
break by fatigue is
\begin{equation}
\delta=\frac{1}{\sum_{j=1}^N k(\sigma_j)},
\end{equation}
Supposing that $\nu_0=1$ and $\sigma_0=1$ in Eq.\ (\ref{eq4015}),
we have
\begin{equation}
\delta=\frac{1}{\sum_{j=1}^N \sigma_j^\rho},
 \label{eq4010}
\end{equation}

In the first step, $\sigma_j=\sigma_i$ $\forall j$ and hence
$\delta=\frac{1}{N\sigma_i^{\rho}}$. This will change with time
because of stress transfers. At any instant of the process of
breaking, $\delta$ of Eq.\ (\ref{eq4010}) represents the mean time
for the next individual failure. The identification of which
element breaks after one $\delta$ is calculated by deciding that
the probability that precisely element $k$ be the affected one is
given by
\begin{equation}
p_k=\sigma_k^{\rho} \delta.
 \label{eq4011}
\end{equation}

The reader will note that, from the probabilistic perspective, one
does not consider weak elements and strong elements: here all
elements are equal but, in general, with a different $\sigma_j$.
The succession of individual breakings proceeds by chance with the
probabilities dictated by Eq.\ (\ref{eq4011}) until the total
collapse of the system. In {\it G\'{o}mez et al.,}
[1998] it was shown that the probabilistic approach represents a
way of partially smoothing the fluctuations inherent to these
stochastic models of fracture.

\section{The model}

The model used in this paper to describe aftershock sequences is
based on the same physical mechanism as the model recently
introduced by {\it Lee} [1999] and {\it Lee and Sornette} [1999],
although the ingredients of the cellular automaton, the way of
running the model and the type of results are different. Our model
is inspired by dynamic FBMs, but with several substantive
differences. As stated above, in dynamic FBMs $N$ uniformly loaded
elements break by fatigue one after one, in a total stress
conserving process until the last individual failure. The sequence
always occurs in a finite-time accelerated process.

Trying to describe a sequence of aftershocks, which is a clearly
decelerated process of relaxation, we will modify three things: ($i$)
stress is lost in the breaking process in two ways, one is by stress
transfers out of the system through the borders, and the other way is
by considering a dissipative effect during each transfer. When an
element fails bearing a stress $\sigma$, the fraction $(1-\pi)\sigma$
is removed from the system. Thus, the constant $\pi$ will represent the
degree of conservation.

($ii$) In FBMs, when an element fails, it transfers its load and
then remains inactive. In this model, on the contrary, when an
element fails, it transfers its load (except the dissipated
fraction) to its nearest neighbors but then it is automatically
regenerated and able to receive stress again and actively
participate in the time evolution of the set. After its
regeneration, an element has zero load before receiving any load.
This assumption is justified because time intervals between
individual failures are much longer than the time of actual
breaking of an asperity, and hence in an interevent interval a
just failed asperity has enough time to reheal.

The third difference ($iii$) with FBMs is related to the initial
stress distribution in the system. Whilst in FBMs it is usually
assumed that the initial load per element is a constant
$\sigma_i$, here, trying to simulate the disordered state existing
in the fault after a main shock, we will take the initial
$\sigma_j$ from a uniform probability distribution ($0 \leq
\sigma_j<1; 1\leq j\leq N$). In the actual running of the model we
will adhere to the probabilistic approach explained above for the
FBM, and so we will deal with the $\delta$s as defined in Eq.\
(\ref{eq4010}) and the probabilities of Eq.\ (\ref{eq4011}).

Besides the three general modifications ($i$), ($ii$), and ($iii$)
introduced with respect to FBMs, in order to reinforce the
appearance of sudden accelerations, which constitute the genuine
phenomenology of the complex aftershocks sequences, we will add
two more rules for running our model.

When a breaking is going to occur in a context such that all the
elements have $\sigma<1$, we say that this is a {\em normal}
event, the $\delta$ is calculated using Eq.\ (\ref{eq4010}) and
the broken element is pointed out by using Eq.\ (\ref{eq4011}). On
the contrary if a breaking is going to occur with one or several
elements with $\sigma\geq 1$, we call it an {\em avalanche} step.
For each avalanche step, {\it a)} the $\delta$ is calculated
using, as usual, Eq.\ (\ref{eq4010}) but the element that breaks
is that one whose $\sigma_j$ surpasses $1$. If there are several
elements with $\sigma >1$, the element that fails is that with the
maximum value of $\sigma_j$. {\it b)} The avalanche ends when all
the $\sigma_j$ of the set become lower than $1$. During an
avalanche, which in general involves several $\delta$s, all the
elements that have surpassed at any step of the avalanche the
condition $\sigma >1$, remain inactive with $\sigma=0$ until the
end of the avalanche.

These two additional rules {\it a)} and {\it b)} are introduced in
order to increase the local stress accumulations. The high stress
concentrations occurring in avalanche events lead to very short
$\delta$s and in very short $\delta$s it is reasonable to assume
that there is not enough time for the healing process.

As a r\'esum\'e of the last paragraphs, we will recall that in the
process of evolution of the system there are normal events and
avalanche events. The former (normal) refers to the failure of one
element when no element in the system has $\sigma>1$. The latter
(avalanche steps) corresponds to the failure of one element with
$\sigma>1$. Eq.\ (\ref{eq4010}) is always used for the calculation
of the time intervals. The $\delta$s of the avalanche steps are
much shorter because of the large stress concentrations induced by
rules {\it a)} and {\it b)}, and the magnitude of the exponent
$\rho$. With these rules, it is obvious that the avalanches become
extinct with time because as the total stress in the system
declines, it is more difficult to locally accumulate load to
surpass unity. The model of {\it Lee} [1999] and {\it Lee and
Sornette} [1999] cannot describe complex aftershock sequences
because in their model the avalanches are instantaneous in time,
so that the decay rate would have singularities at the time of an
avalanche. On the contrary, our avalanches are not instantaneous
since they are formed by several steps with their corresponding
$\delta$s, that is, they occur in a finite time interval.

We have performed our simulations in a two dimensional square
lattice with $50x50$ sites with $\rho=30$ and $\pi=0.7$. Each site
represents the emplacement of one element or asperity. The
load-transfer rule assumed is a local load-sharing (LLS) rule: a
failing element transfers its load to its nearest four neighbors
located in the North, South, East and West. If the element is
located at the borders of the square lattice this isotropic
transfer provokes the corresponding stress leakage.

In FBMs, in which broken elements remain inactive, an LLS rule can
lead to an uncertain situation when a failing element lacks active
neighbors able to receive the stress transfer. In this model,
though, this does not occur because any element, at any time, is
active to receive stress. The only possible exceptional situation
could come from the application of rule {\it b)} in avalanches.
Then, the load of the failing element is transferred only to the
existing active nearest neighbors. In this extremely rare case (we
have not met such a situation in our simulations) of having the
four nearest neighbors already broken, the load is removed from
the lattice. This is just one possibility among various choices.
 {\it Lomnitz-Adler et al.} [1992] explored in detail
three of these possibilities for a cellular automaton with rules
similar to those of the LLS static fiber-bundle model, and the
reader is referred to this paper for details. In our model
deciding among one of the three alternative scenarios is not
important due to the extreme rarity of such events.

The running of the program proceeds as follows. At $t=0$ all the
elements on the lattice are initialized by loading them with a
random initial stress $\sigma_j$ taken from a uniform probability
distribution $0\leq \sigma_j \leq 1$. Then, we calculate the
$\delta$ corresponding to the first failure by using Eq.\
(\ref{eq4010}). The choice of the element that actually breaks is
done by using Eq.\ (\ref{eq4011}), and a random number between 0
and 1 to materialize the choice. The chosen element fails and
$\pi$ times its stress is transferred to its nearest neighbors and
the $(1-\pi)$ fraction disappears. Now, we analyze the
distribution of stress in the board; if all the $\sigma_i$ are
lower than one, the process of calculating the next failure is
identical to the first. If, on the contrary, one (or several)
$\sigma_j
>1$ then we calculate the corresponding value of $\delta$ from Eq.\
(\ref{eq4010}) and the failure is assigned to the element with the
biggest $\sigma_j >1$. During the period of an avalanche, rule {\it b)}
is applied to favor the stress concentration. Thus, the series of
breakings and transfers, involving normal events or avalanches,
proceeds until a prescribed minimum value for the total stress on the
system is reached, otherwise there will be an infinite number of
aftershocks.

In our model, due to the dissipation, the total stress in the system,
$S=\sum_{j}\sigma_{j}$, systematically decreases. If the value $\rho=1$
were considered, then from Eq.\ (\ref{eq4010}), the successive
$\delta$s would necessarily be longer and longer. But $\rho$ is bigger
than one, and this is the reason why one can have a step down in $S$
and find a shorter value of $\delta$. This is the key point to
understand our model and other models based on subcritical crack
growth. In the general trend of $S$ reduction and hence temporal
deceleration, the stress transfers in the system provoke local
inhomogeneities in $\sigma$, and due to the high values of $\rho$, this
leads to temporal accelerations. These accelerations are embedded in
the general trend of dynamical relaxation.

\section{Results and Conclusions}

We have carried out numerical simulations which show the
fulfillment of Omori's law and reproduce the features already
commented on, that is, a cumulative plot with sudden variations in
the number of events (accelerations). We show here the results for
a two-dimensional system of $50\times50$ elements located on a
square lattice, with $\rho$ equal to $30$ and a conservation level
of $\pi=0.7$. Other simulations have also been performed varying
the size of the system, the value of $\rho$ and the conservation
level, $\pi$. We have also explored a variant of rule $b)$ in
which, in avalanches, all the elements with $\sigma\geq 1$ break
simultaneously in the same $\delta$. We have found that the
results are indeed very close to those exposed here with equal
qualitative behavior. Nevertheless, it should be noted that
although the results are robust over a large range of parameters,
different characteristics arise for extreme values of $\rho$ and
$\pi$.

Figure \ref{fig:mrate} shows the rate of aftershocks $dN/dt$ as a
function of time. Time is represented in dimensionless units and
it is the sum of the successive $\delta$s. The straight line has a
slope of $-1$ for comparison. Thus, the $1/t$ decay is confirmed
and is in full agreement with Omori's law for real aftershock
sequences. The power law depicted is very robust over a wide range
of the parameters that characterize the model. The most critical
parameter is the conservation level since for values of $\pi$
close to unity, the system does not dissipate enough to avoid its
complete failure. Besides, for large dissipation $\pi\ll1$, the
power law extends only to a few decades, and the number of decades
decreases as $\pi$ decreases. Anyhow, in all cases the exponent of
the power law decay is very close to unity. The major vertical
spikes of Fig.~\ref{fig:mrate} correspond to avalanche-type events
that disappear with time. The smaller fluctuations for large times
reflect the intrinsic probabilistic nature of the model and are
not related at all to the appearance of avalanches. This is
clearly seen in Fig.~\ref{fig:macum} where we have plotted the
cumulative number of aftershocks versus time instead of the
differential plot of Fig.~\ref{fig:mrate}. As can be observed in
Fig.~\ref{fig:macum}, sudden accelerations appear in the first
stages of rupture. This behavior closely resembles that previously
reported in Sec. 2 for the eastern Pyrenees aftershock sequence
(see Fig.~\ref{fig:acum}). The Omori's law maximum likelihood fit
to the model results is also included in Fig.~\ref{fig:macum}. The
estimated parameter values are: $p=1.01 \pm 0.06$, $K=210 \pm 29 $
and $c=1.3 \pm 0.2$.

In our model, the changes in aftershock rate are related to the
readjustments of local stresses when events take place. With time,
local concentrations of stress appear in the system and there is a
high probability of finding a region in which the load supported
by the elements is close to the threshold value $\sigma=1$. That
is, there is a large heterogeneous stress state in which one
failure will trigger an avalanche. During the evolution of the
avalanche the local accumulation of load increases. This fact
together with the high value of $\rho$, provokes the $\delta$s
corresponding to these steps of rupture to be considerably
reduced. As a result, we observe the step-like change in the
cumulative number of events (contrast Fig.5 with Fig.6). For large
times the avalanches eventually disappear since in a
non-conservative model the total load in the system systematically
decreases and hence it would be unlikely to accumulate stress in
local regions as to surpass the value $\sigma=1$.

Of interest is the further investigation of the acceleration
events in order to get an additional insight about the observed
aftershock sequences. One simple way to do that was explained in
Sec.\ 2. It consists of decomposing the original series of
aftershocks in leading events and cascades depending on whether a
relaxation law is accomplished or not. We have followed the same
procedure with the synthetic data. The series of cascades obtained
in such a way is shown in Fig.~\ref{fig:mcascades}. Part (a) shows
the series after removing all the cascades, i.e., leaving only
leading aftershocks, and in Fig.~\ref{fig:mcascades}b we plot the
cascades, in which the first event of each cascade is a leading
aftershock. The decomposition obtained from the model is indeed
indistinguishable from that corresponding to the real series of
events (Fig.~\ref{fig:cascades}). Two characteristics of the
series of cascades are again relevant: one, the elapsed time
between successive leading events is larger than the preceding one
in complete agreement with the supposition of a relaxation
process, and second, the series of cascades can be well
approximated by almost straight segments whose slopes decrease as
time passes. This later characteristic could be used to quantify
the observed jumps in the cumulative plot of aftershocks and to
explain why they are present mainly in the first stages of
rupture. The larger jumps are related to the occurrence of
avalanches, which are caused by local accumulations of stress; so
it is expected that when avalanches damp out due to dissipation,
changes in the rate of occurrence are more spaced in time and
cascades consist of fewer events. Of course, there will be
fluctuations about the power law trend even in the case where
avalanches have deceased. Thus, we expect slope values gradually
closer to zero as time tends to infinity. This is clearly
appreciated in Fig.~\ref{fig:mslopes}, where we have represented
in a log-log plot the slopes of the cascades versus the occurrence
time of the leading event that initiate each cascade. Plot (a)
shows only the first part of the time sequence to facilitate
comparison with Fig.~\ref{fig:slopes}.  As for the observed series
of aftershocks, the slopes fit a power law with an exponent of
about $\nu=0.94$ for the fist part of the time series and
$\nu=1.08$ for the long-tail end (plot (b)). Actually, the
long-tail exponent $\nu$ ranges between $1.00$ and $1.08$
depending on the conservation level $\pi$ and the value of $\rho$.
This appears to be a smooth dependence. Thus, the qualitative
behavior is again captured. The discrepancy between the slopes
($\nu=0.7$ for the Pyrenees sequence) is not surprising due to the
simplicity of the model as compared with the inherent complexity
of the real phenomenon we want to simulate. The reason for this
particular behavior, that is, why the slopes follow a power law
and no other law, is unclear to us up to now.

Future efforts will be devoted to the understanding of other
dynamical characteristics of the model and their fine dependence
on $\rho$ and $\pi$ by studying another complex series of
aftershocks. We also plan to perform a detailed analysis of the
spatial structure of the sequence of events coming from our model.

\section{ACKNOWLEDGMENTS} Y.M thanks the AECI for financial support. This work
was supported in part by the Spanish DGYCYT (Project PB98-1594).

\section*{REFERENCES}

\begin{description}
\item Atkinson, B.K., Subcritical crack growth in geological
materials, {\it J. Geophys. Res., B89}, 4077-114, 1984.
\item Atkinson, B.K., and P.G. Meredith, The theory of
subcritical crack growth with application to minerals and rocks,
In B.K. Atkinson (Ed), {\it Fracture Mechanics of Rocks} (Academic
Press, London), 111-166, 1987.
\item Barnet R.L., and R. Kerrich, Stress corrosion cracking
of biotite and feldspar, {\it Nature, 283}, 185-187, 1980.
\item Benioff, H., Earthquakes and rock creep, {\it Bull.
Seism. Soc. Am., 41}, 31-62, 1951.
\item Bonn, D., H. Kellay, M. Prochnow, K. Ben-Djemiaa, and
J. Meunier, Delayed fracture of an inhomogeneous soft solid, {\it
Science, 280}, 265-267, 1998.
\item Brace, W.F., and J.D. Byerlee, Stick-slip as a
mechanism of earthquakes, {\it Science, 153}, 990-992, 1966.
\item Burridge, R., and Knopoff, L., Model and theoretical
seismicity, {\it Bull. Seism. Soc. Am., 57}, 341-371, 1967.
\item Conchard, A., and R. Madariaga, Complexity of
seismicity due to highly rate-dependent friction, {\it J. Geophys.
Res., B101}, 25321-36, 1996.
\item Coleman, B.D., Time dependence of mechanical breakdown
in bundles of fibers, {\it J. Appl. Phys. 28}, 1058, 1957.
\item Coleman, B.D., Time dependence of mechanical breakdown
in bundles of fibers, III: The power law breaking rule, {\it
Trans. Soc. Rheology, 2}, 195-218, 1958.
\item Correig, A.M., M. Urquiz\'u, J. Vila, and S. Manrubia,
Aftershock series of event February 18, 1996: An interpretation in
terms of self-organized criticality, {\it J. Geophys. Res., B102},
27,407- 20, 1997.
\item Daniels, H.E., The statistical theory of the strength
of bundles of threads {\it Proc. Roy. Soc. Lond. A183}, 404, 1945.
\item Das, S., and C.H. Scholz, Theory of time-dependent
rupture in the earth, {\it J. Geophys. Res., B86}, 6039-51, 1981.
\item Dieterich, J.H., Time-dependent friction in rocks, {\it
J. Geophys. Res., 77}, 3690-97, 1972a.
\item Dieterich, J.H., Time-dependent friction as a possible
mechanism for aftershocks, {\it J. Geophys. Res., 77}, 3771-81,
1972b.
\item Dieterich, J.H., Modeling of rock friction, 1,
Experimental results and constitutive equations, {\it J. Geophys.
Res., B84}, 2161-68, 1979.
\item Dieterich, J.H., A constitutive law for rate of
earthquake production and its application to earthquake
clustering, {\it J. Geophys. Res., B99}, 2601-18, 1994.
\item Garcimart\'{\i}n, A., A. Guarino, L. Bellon, and S.
Ciliberto, Statistical properties of fracture precursors, {\it
Phys. Rev. Lett., 79}, 17, 3202-05, 1997.
\item Gibowicz, S.J., An anatomy of a seismic sequence in a
deep gold mine, {\it PAGEOPH, 150}, 393-414, 1997.
\item G\'omez, J.B., Y. Moreno, and A.F. Pacheco,
Probabilistic approach to time-dependent load-tranfer models of
fracture, {\it Phys. Rev. E58}, 1528-32, 1998.
\item Gross, S. J., Kisslinger, C., Tests of Models of Aftershock
Decay Rate, {\it Bull. Seismol. Soc. Am., 84}, 1571- 1579, 1994.
\item Gu, J.C., J.R. Rice, A.L. Ruina, and S.T. Tse, Slip
motion and stability of a single degree of freedom elastic system
with rate and state dependent friction, {\it J. Mech. Phys.
Solids, 32}, 167-196, 1984.
\item Heaton, T.H., Evidence for and implications of
self-healing pulses of slip in earthquake rupture, {\it Phys.
Earth Planet. Inter., 64}, 1-20, 1990.
\item Hirata, T., Omori's power law aftershock sequences of
microfracturing in rock fracture experiments, {\it J. Geophys.
Res., B92}, 6215-21, 1987.
\item Jacobs, D.S., and I.W. Chen, Mechanical and
environmental factors in the cyclic and static fatigue of
silicon-nitride, {\it J. Am. Ceram. Soc., 77}, 1153-1161, 1994.
\item Kostrov, B.V., L.V. Nikitin, and L.M. Flitman, The
mechanics of brittle fracture, {\it Mech. Solids, 3}, 105-117,
1969.
\item Lee, M.W., {\it Unstable fault interactions and
earthquake self-organization}, PhD Thesis, UCLA, 1999.
\item Lee, M.W., and D. Sornette, Novel mechanism for
discrete scale invariance in sandpile models, {\it
cond-mat/9903402}, 27 Mar 1999.
\item Lockner, D., The role of acoustic emission in the study
of rock, {\it Int. J. Rock Mech. Min. Sci. \& Geomech. Abstr.,
30}, 883-899, 1993.
\item Lockner, D., and Byerlee, J.D., Acoustic emission and
creep in rocks at high confining pressures and differential
stress, {\it Bull. Seism. Soc. Am., 67}, 247-258, 1977.
\item Lomnitz-Adler, J., Knopoff, L. and
Mart\'{\i}nez-Mekler, G., Avalanches and epidemic models of
fracturing in earthquakes, {\it Phys. Rev. A45}, 2211-2221, 1992.
\item Marcellini, A., Arrhenius behavior of aftershock
sequences, {\it J. Geophys. Res., B100}, 6463-68, 1995.
\item Marcellini, A., Physical model of aftershock temporal
behavior, {\it Tectonophysics, 277}, 137- 146, 1997.
\item Newman, W.I., D.L. Turcotte, and A.M. Gabrielov,
Log-periodic behavior of a hierarchical failure model with
applications to precursory sesimic activation, {\it Phys. Rev.
E52}, 4827-35, 1995.
\item Ogata, Y., Statistical models for earthquake occurrence
and residual analysis for point processes, {\it J. Am. Stat.
Assoc., 83}, 9-27, 1988.
\item Okoroafor, E.U., and R. Hill, Investigations of complex
failure modes in fibre bundles during dynamic mechanical testing
using acoustic emission and Weibull statistics, {\it J. Materials
Sci., 30}, 4233-43, 1995.
\item Omori, F, On after-shocks of earthquakes, {\it J. Coll.
Sci. Imp. Univ. Tokyo, 7}, 111-200, 1894.
\item Pauchard, L, and J. Meunier, Instantaneous and time-lag
breaking of a two-dimensional solid rod under a bending stress,
{\it Phys. Rev. Lett., 70}, 23, 3565-68, 1993.
\item Pauchet, H.A., A. Rigo, L. Rivera, A. Sourian, A
detailed analysis of the February 1996 aftershock sequence in the
eastern Pyrenees, France, {\it Geophys. J. Inter., 137}, 107-127,
1999.
\item Phoenix, S.L., Stochastic strength and fatigue in fiber
bundles, {\it Int. J. Fracture, 14}, 327-344, 1977.
\item Phoenix, S.L., and L. Tierney, A statistical model for
the time dependent failure of unidirectional compositie materials
under local elastic load-sharing among fibers, {\it Eng. Frature
Mech., 18}, 193-215, 1983.
\item Rice, J.R., and A.L. Ruina, Stability of steady
frictional slipping, {\it J. Appl. Mech., 50}, 343-349, 1983.
\item Rice, J.R., and S.T. Tse, Dynamic motion of a single
degree-of-freedom system  following a rate- and state-dependent
friction law, {\it J. Geophys. Res., B91}, 521-530, 1986.
\item Rigo, A., C. Olivera, A. Sourian, S. Figueras, H.
Paucher, A. Gr\'esillaud, and M. Nicolas, The February 1996
earthquake sequence in the eastern Pyrenees: first results, {\it
J. Seismology, 1}, 3-14, 1997.
\item Ruina, A.L., Slip instability and state variable
friction laws, {\it J. Geophys. Res., B88}, 10,359-70, 1983.
\item Sammonds, P.R., P.G. Meredith, and I.G. Main, Role of
pore fluids in the generation of sesimic precursors to shear
fracture, {\it Nature, 359}, 228-230, 1992.
\item Schleinkofer, U., H.G. Sockel, K. Gorting, and W.
Heinrich, Fatigue of hard metals and cermets, {\it Mat. Sci. Eng.
A, 209}, 313-317, 1996.
\item Scholz, C.H., The frequency-magnitude relation of
microfracturing in rock and its relation to earthquakes, {\it
Bull. Seismol. Soc. Am., 58}, 399-415, 1968a.
\item Scholz, C.H., Microfractures, aftershocks, and
seismicity, {\it Bull. Seismol. Soc. Am., 58}, 1117- 30, 1968b.
\item Scholz, C.H., {\it The Mechanics of Earthquakes and
Faulting}, Cambridge University Press, New York, USA, 1990.
\item Talebi, S. (ed), Seismicity associated with mines, reservoirs and fluid
injections, {\it PAGEOPH, 150}, 379-720, 1997.
\item Utsu, T., A statistical study on the occurrence of
aftershocks, {\it Geophys. Mag., 30}, 521-605, 1961.
\item Utsu, T, Y. Ogata, and S Matsu'ura, The centenary of
the Omori formula for a decay law of aftershock activity, {\it J.
Phys. Earth, 43}, 1-33, 1995.
\item Vazquez-Prada, M, J.B. G\'omez, Y. Moreno, and A.F.
Pacheco, Time to failure of hierarchical load-transfer models of
fracture, {\it Phys. Rev. E}, in press, 1999.
\item Wennerberg, L., and Sharp, R.V., Bulk-friction of
afterslip and the modified Omori law, {\it Tectonophysics, 277},
109-136, 1997.
\item Wiederhorn, S.M., Influence of water vapor on crack
propagation in soda-lime glass, {\it J. Amer. Ceram. Soc, 50},
407-418, 1967.
\item Yamashita, T, and L. Knopoff, Model of aftershock
occurrence, {\it Geophys. J. R. Astron. Soc., 91}, 13-26, 1987.
\item Zhurkov, S.N., Kinetic concept of the strength of
solids, {\it Int. J. Fracture Mech., 1}, 311-323, 1965.
\end{description}

\newpage

\begin{figure}
\caption[]{\label{fig:acum} Aftershock sequence of February 18,
1996, eastern Pyrenees. (a) Complete series of aftershocks, shown
as the accumulated number of events (left axis), together with
their magnitude (right axis). (b) First 300 hours of the
aftershock sequence, as used in the comparison with the model
results. The fit to the whole range and two independent fits to
the 0-100 h and 140-300 h intervals has been superimposed (c)
Separation of the aftershock sequence into leading aftershocks
(filled circles) and cascade events (dots). See the text for
details. }
\end{figure}

\begin{figure}
\caption[]{\label{fig:cascades}(a) Series formed by the leading
aftershocks, after removal from the original sequence the
cascades. The fit to the Omori law is much better than the
original, and the $p$ value (0.94) is also closer to worldwide
aftershock $p$-values. (b) Cascades retrieved from the original
first 300 hours of the aftershock sequence. Each cascade can be
approximated by a straight line.}
\end{figure}

\begin{figure}
\caption[]{\label{fig:slopes} Slope of the cascades versus time
on log-log scale. It can be clearly seen that it follows a power
law $s\propto t^{-\nu}$, with $\nu \approx 0.7$ }
\end{figure}

\begin{figure}
\caption[]{\label{fig:mrate} Rate of aftershocks $dN/dt$ as a
function of dimensionless time for a dissipation of $\pi=0.7$ and
a Weibull index of $\rho=30$. The spikes that decorate the general
$t^{-1}$ trend correspond to sudden accelerations in event rate
(avalanches). The diagonal straight line has a slope of $-1$. This
curve was obtained by numerically differentiating the curve in
Fig. \ref{fig:macum}  }
\end{figure}

\begin{figure}
\caption[]{\label{fig:macum} Accumulated number of aftershocks $N$
as a function of dimensionless time for a dissipation of $\pi=0.7$
and a Weibull index of $\rho=30$. Note the sudden increases in
event rate (step-like jumps) superimposed to the general Omori-law
trend. The continuous line is a maximum likelihood fit to the 2-10
range and the corresponding {\it p}-value is $p=1.01\pm 0.06$. }
\end{figure}

\begin{figure}
\caption[]{\label{fig:mcascades} (a) Leading aftershock sequence
for a simulation with $\pi=0.7$ and $\rho=30$. (b)Model cascades.
The first event in each cascade is a leading aftershock. Note that
the cascades can be also approximated by straight lines, as was
the case with the cascades in the actual aftershock sequence,
Fig.\ 2. }
\end{figure}

\begin{figure}
\caption[]{\label{fig:mslopes} (a) Slopes of the model cascades
versus dimensionless time of the leading event that initiates each
cascade on log-log scale. Only the first part of the model
cascades is shown in this plot to facilitate comparison with Fig.\
3. The local power-law exponent in this part is $\nu=0.94$. (b)
Log-log representation of the slopes versus time for the long-time
tail of the simulation. As for the eastern Pyrenees series of
aftershocks, the slopes fit very well a power law, in this case
with an exponent of about $1.08$. }
\end{figure}

\end{multicols}
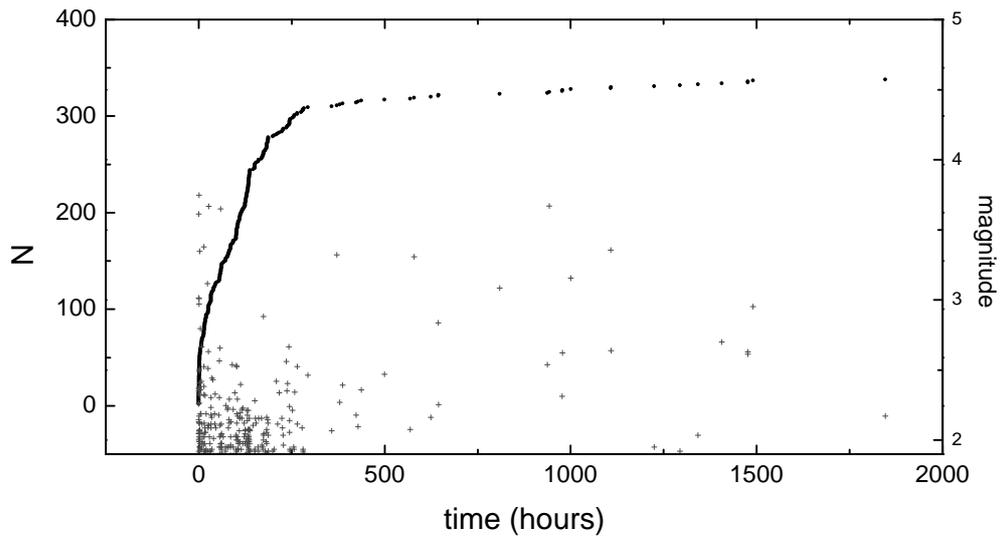
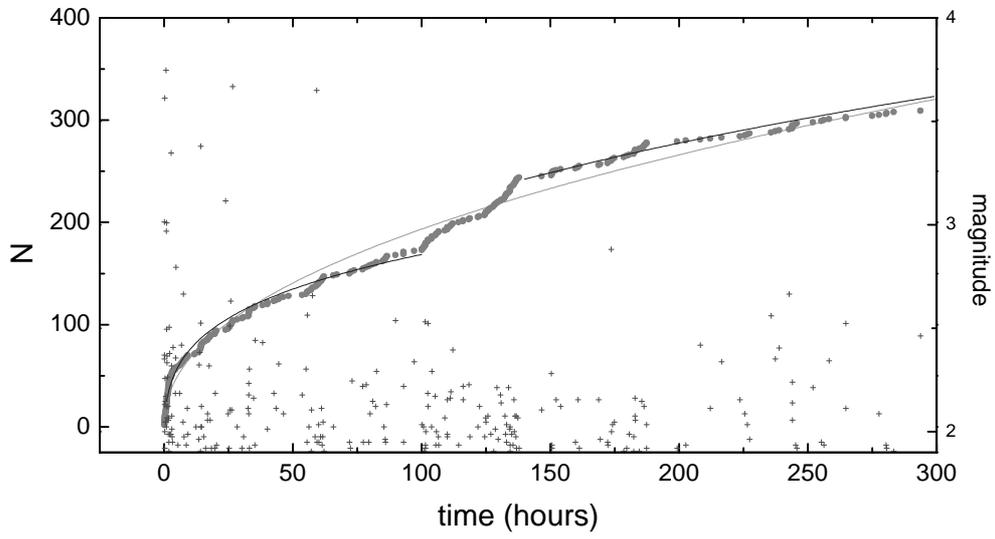
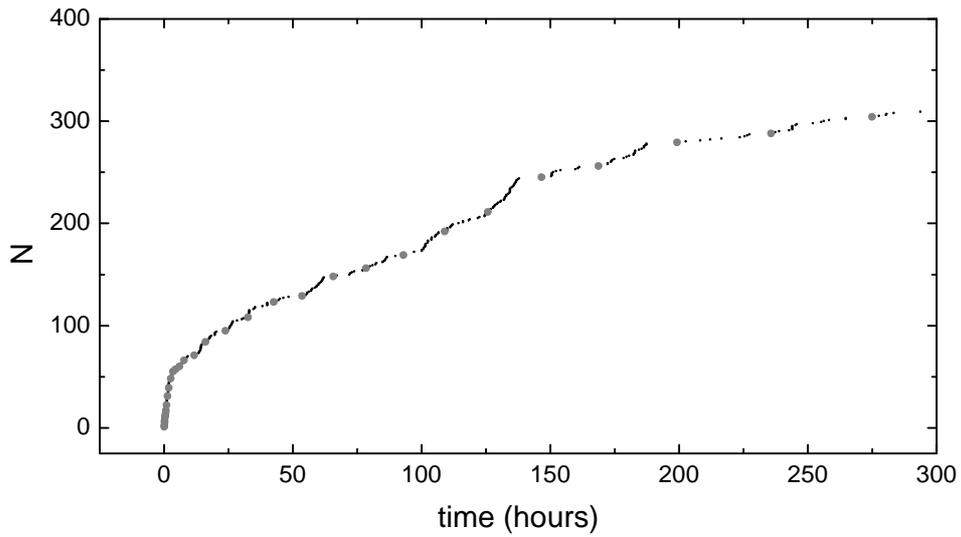
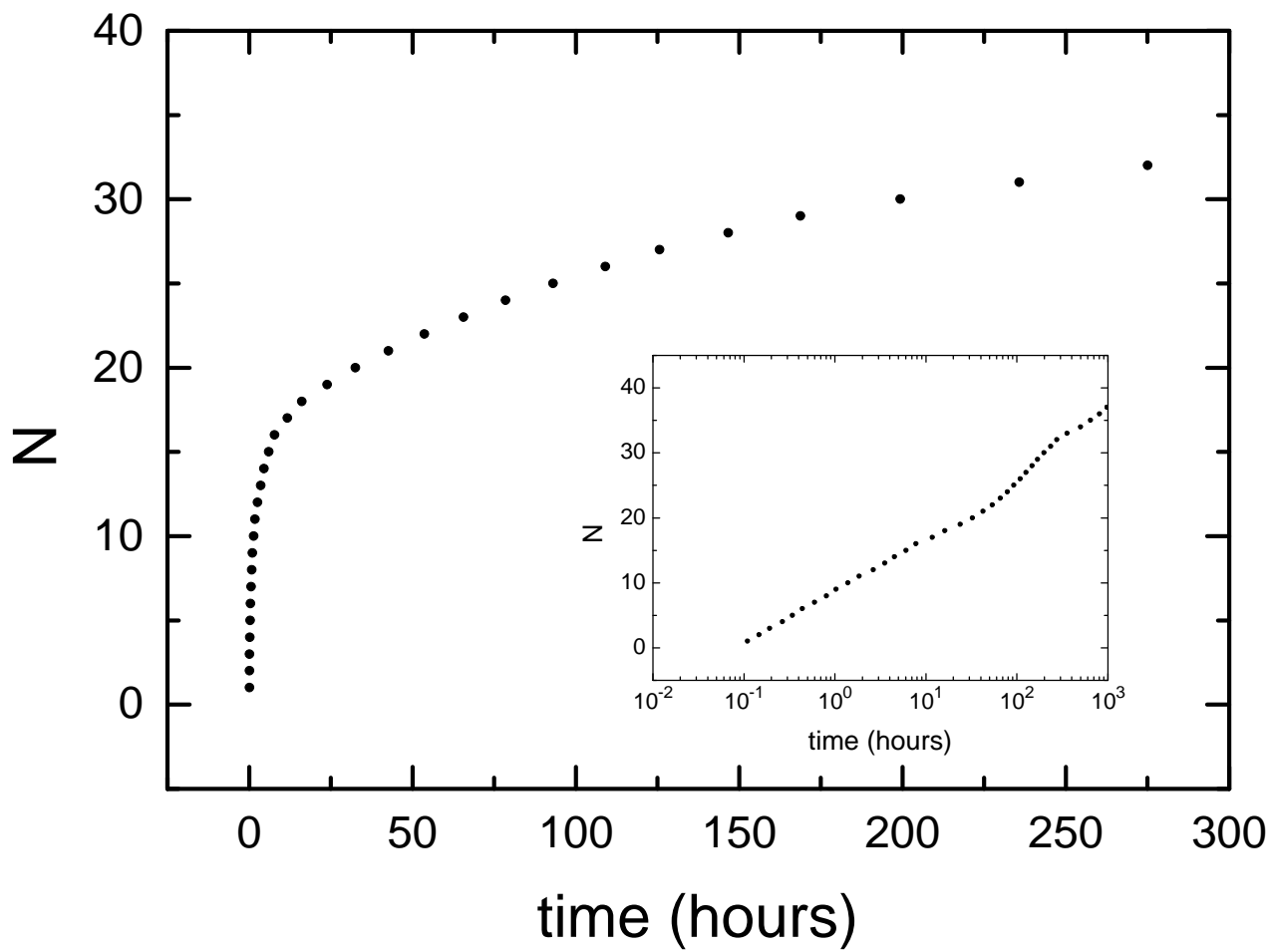
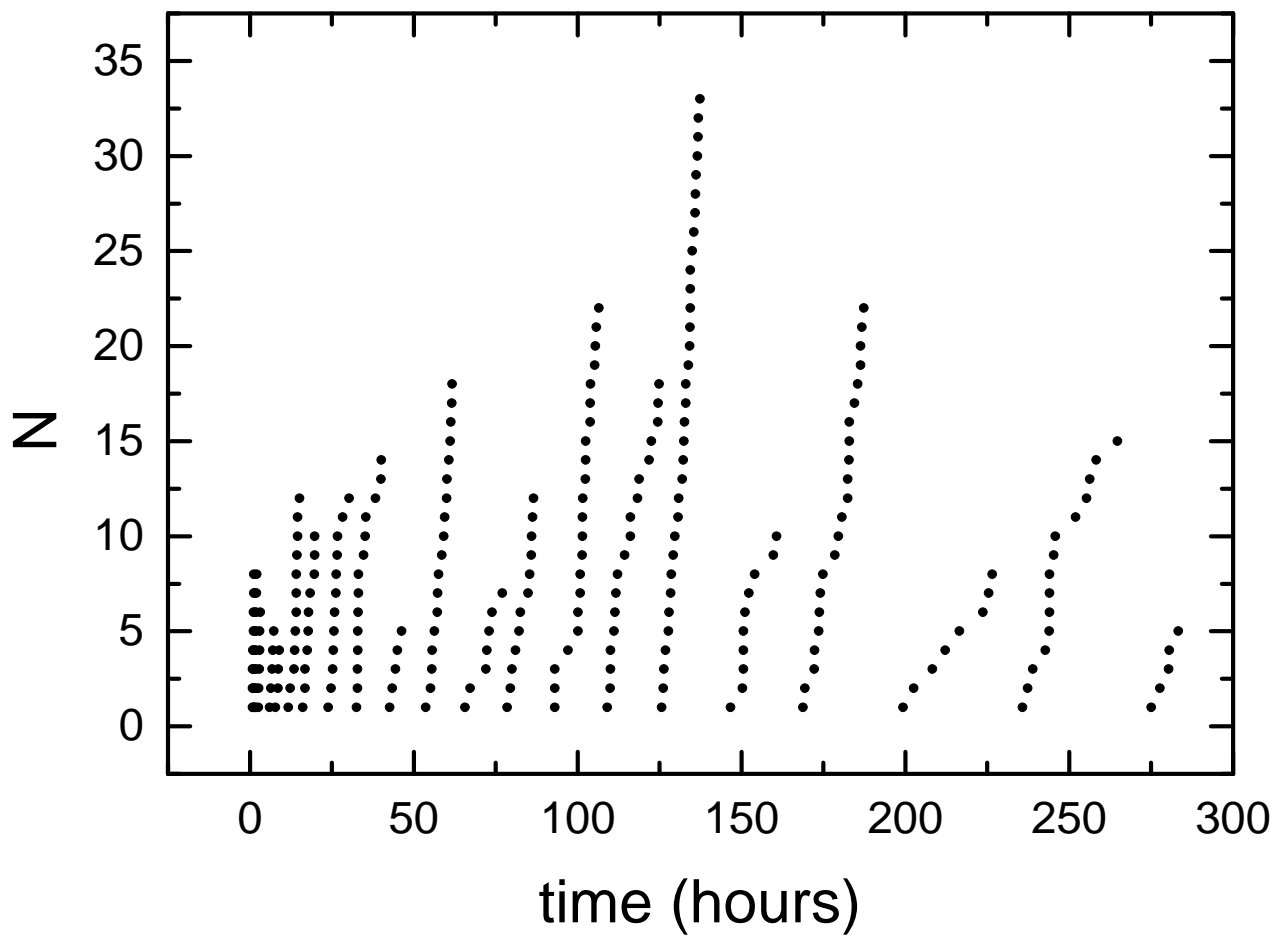
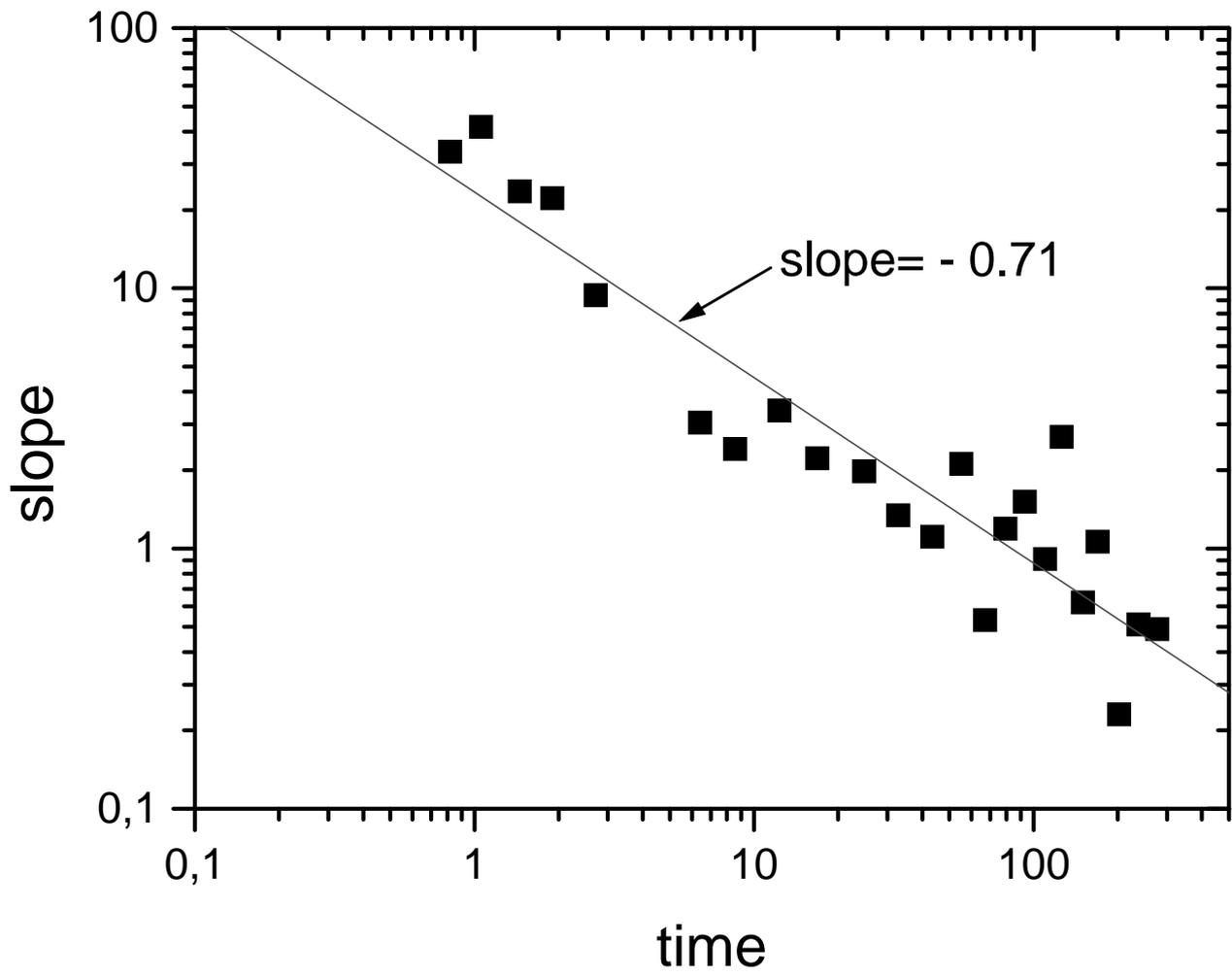
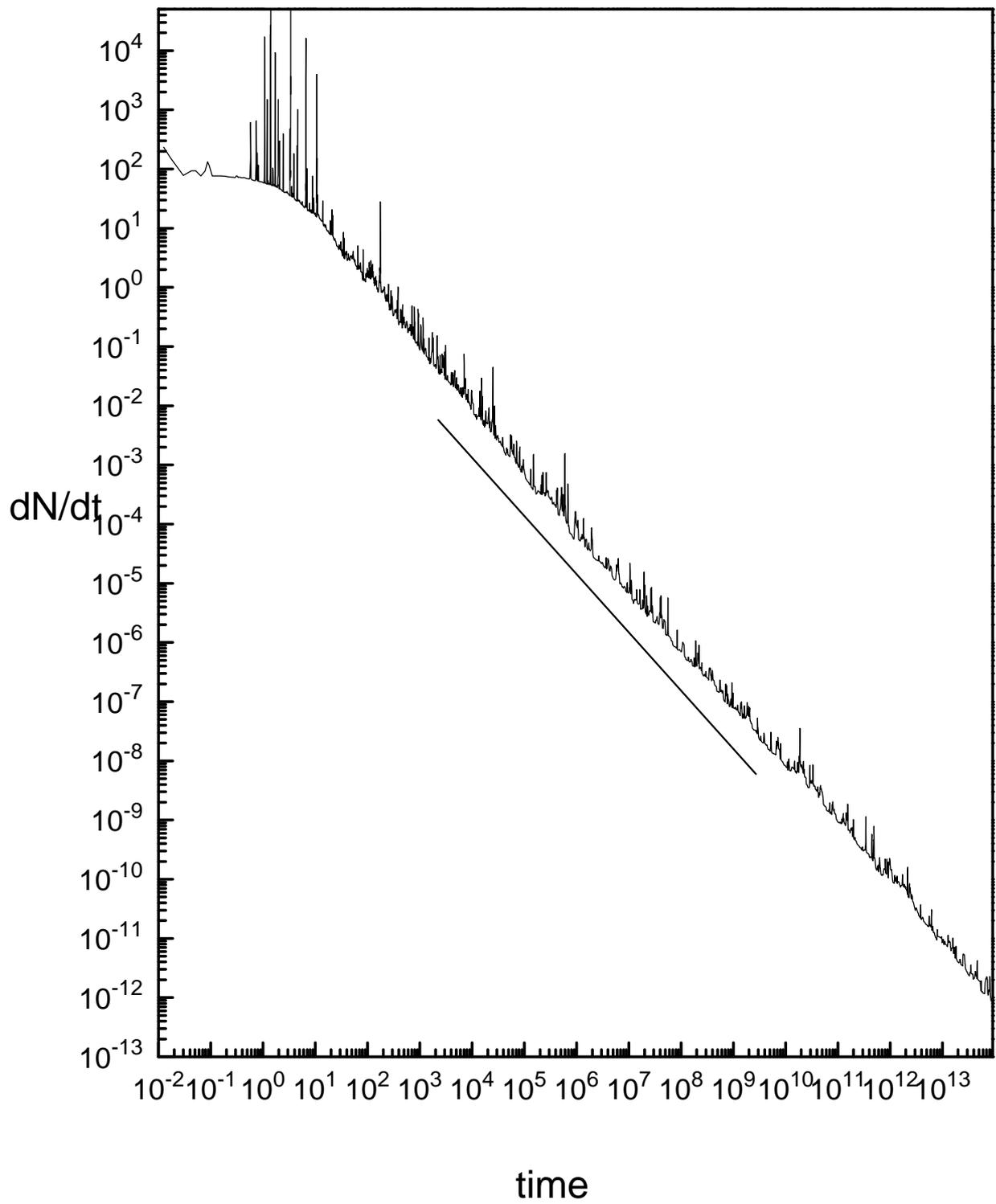
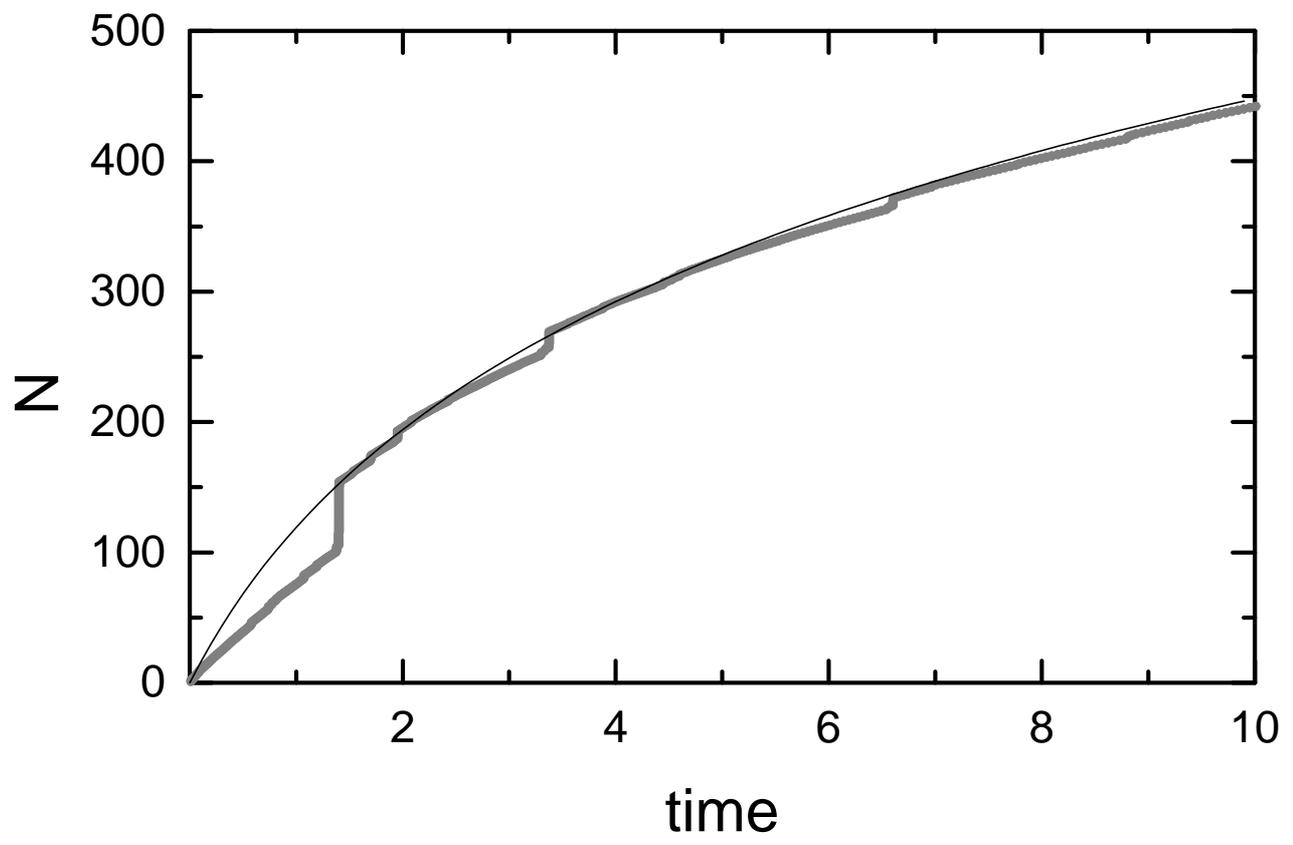
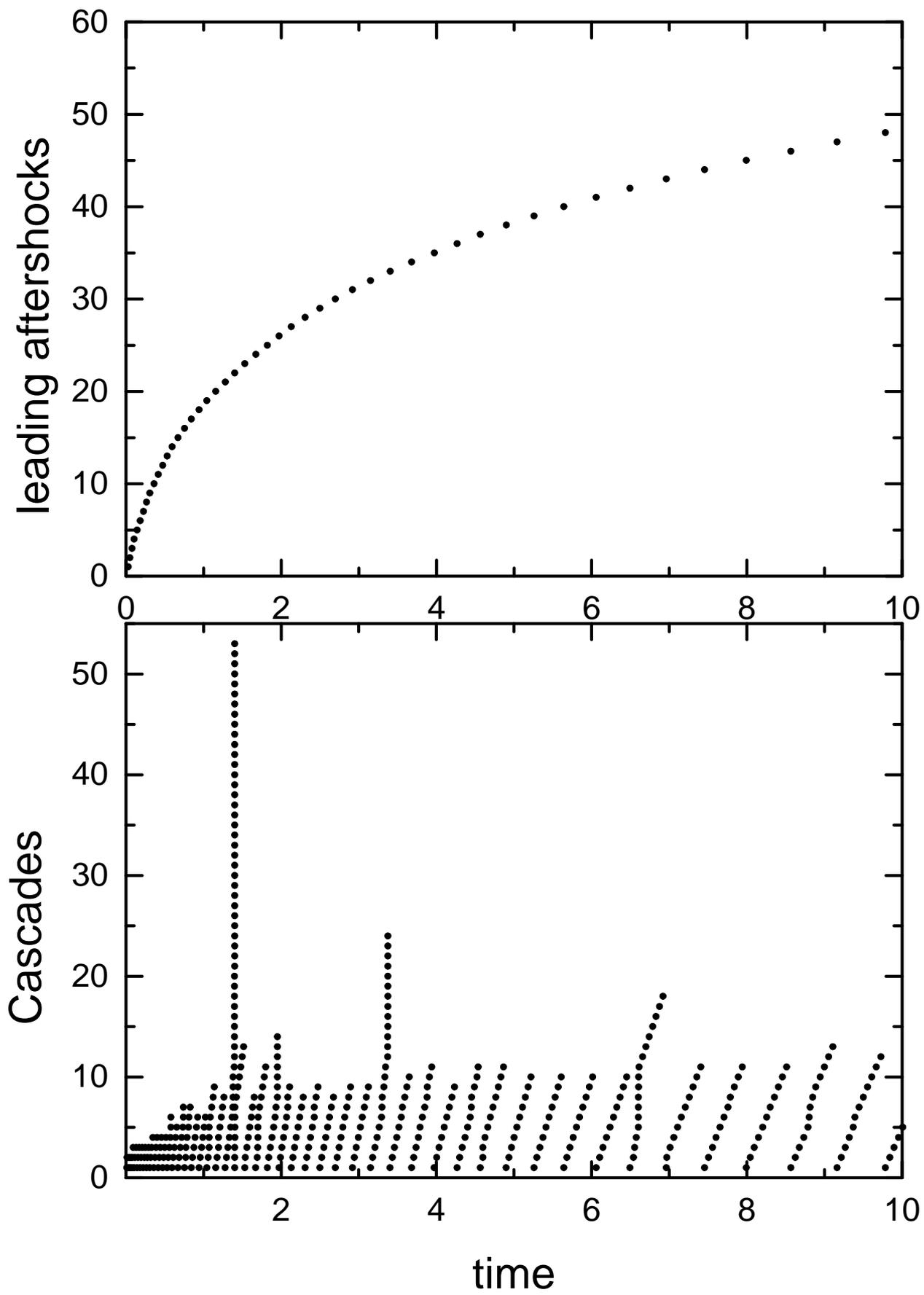
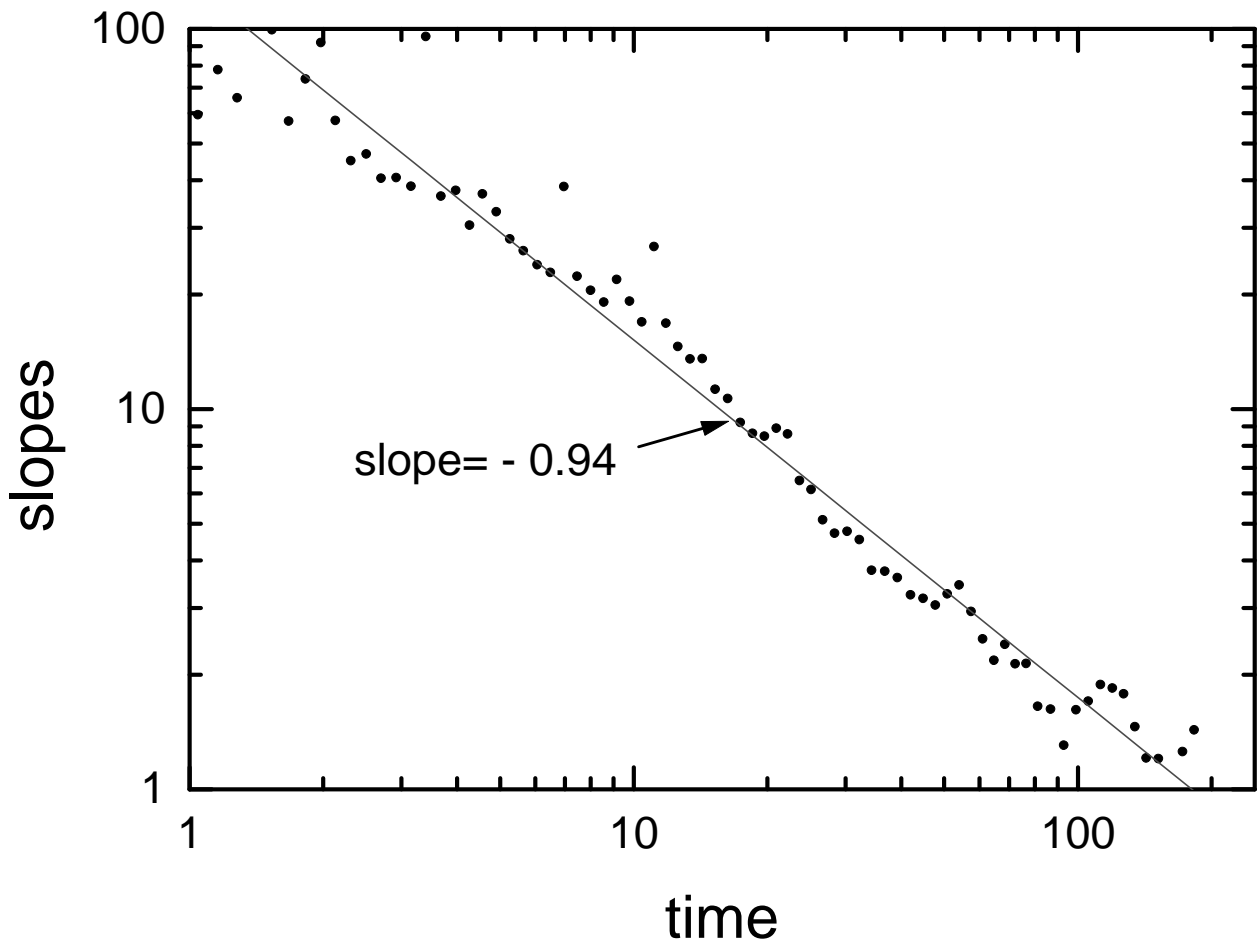
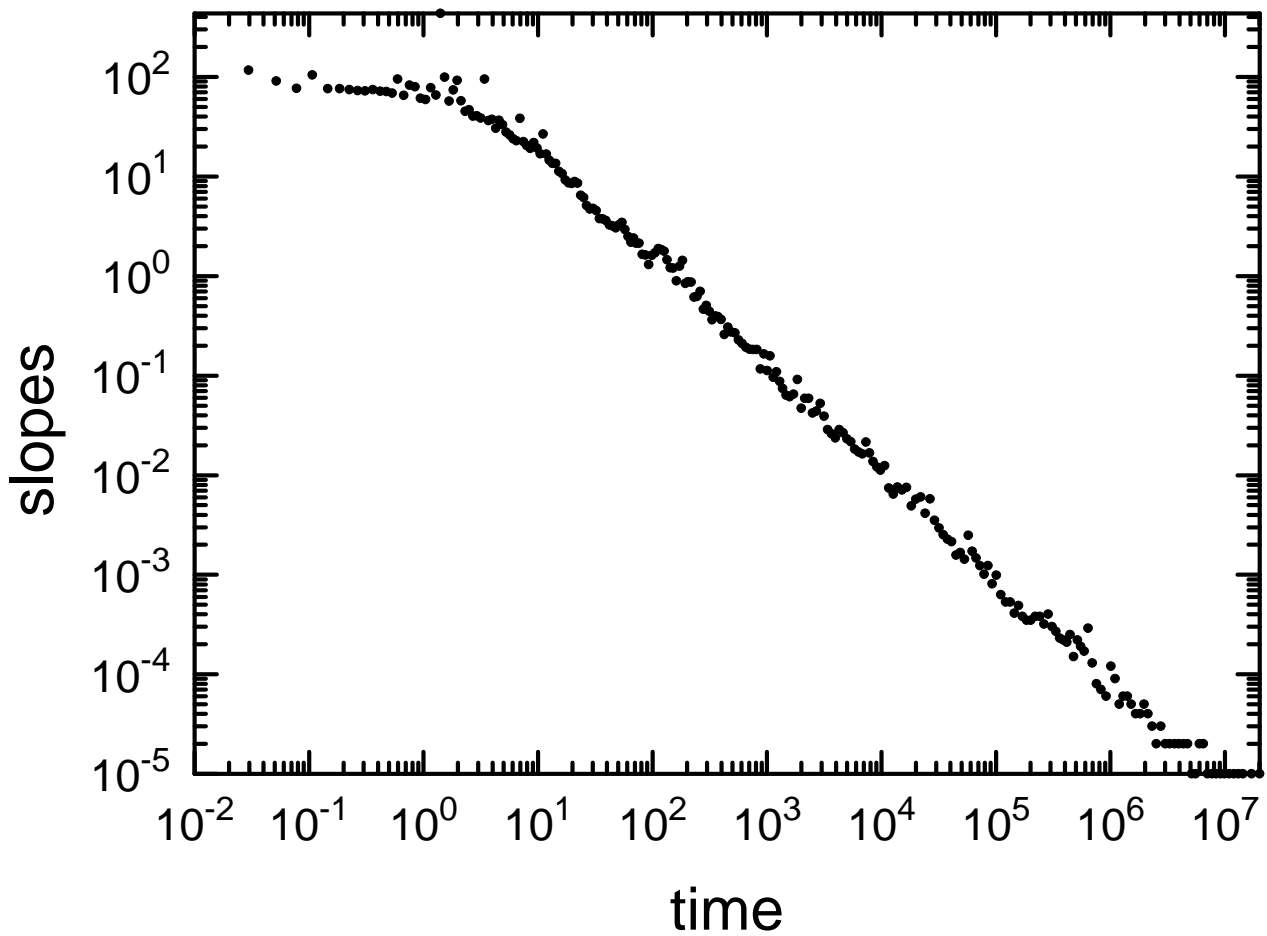
\end{document}